\documentclass{aa}  
\usepackage{natbib}
\usepackage{color}
\usepackage{graphicx}
\usepackage{txfonts}
 \def\mean#1{\left< #1 \right>}
 \def\SF{S\negthinspace F}

\begin{document} 

   \title{Ensemble X-ray variability of active galactic nuclei. \\ II. Excess variance and updated structure function\thanks{Table 1 is only available in electronic form
at the CDS via anonymous ftp to cdsarc.u-strasbg.fr (130.79.128.5) or via http://cdsweb.u-strasbg.fr/}}

   \author{F. Vagnetti \inst{1}, R. Middei \inst{1,2}, M. Antonucci \inst{1}, M. Paolillo \inst{3}, \and R. Serafinelli \inst{1}}
   
   \offprints{F. Vagnetti}

   \institute
       {Dipartimento di Fisica, Universit\`a di Roma ``Tor Vergata'', Via della Ricerca Scientifica 1, I-00133, Roma, Italy   \\ \email{fausto.vagnetti@roma2.infn.it}
      \and
       Dipartimento di Matematica e Fisica, Universit\`a Roma Tre, via della Vasca Navale 84, 00146 Roma, Italy\thanks{Present address}
      \and
       Dipartimento di Scienze Fisiche, Universit\`a di Napoli Federico II, Via Cinthia 9, 80126 Napoli, Italy}

   \date{}

 
  \abstract
  {Most investigations of the X-ray variability of active galactic nuclei (AGN) have been concentrated on the detailed analyses of individual, nearby sources. A relatively small number of studies have treated the ensemble behaviour of the more general AGN population in wider regions of the luminosity-redshift plane.} 
  {We want to determine the ensemble variability properties of a rich AGN sample, called Multi-Epoch XMM Serendipitous AGN Sample (MEXSAS), extracted from the fifth release of the XMM-Newton Serendipitous Source Catalogue (XMMSSC-DR5), with redshift between $\sim 0.1$ and $\sim 5$, and X-ray luminosities in the 0.5-4.5 keV band between $\sim 10^{42}$ erg/s and $\sim 10^{47}$ erg/s.}
  {We urge caution on the use of the normalised excess variance (NXS), noting that it may lead to underestimate variability if used improperly. We use the structure function (SF), updating our previous analysis for a smaller sample. We propose a correction to the NXS variability estimator, taking account of the light curve duration in the rest frame on the basis of the knowledge of the variability behaviour gained by SF studies.}
  {We find an ensemble increase of the X-ray variability with the rest-frame time lag $\tau$, given by $\SF\propto\tau^{0.12}$. We confirm an inverse dependence on the X-ray luminosity, approximately as $\SF\propto L_X^{-0.19}$. We analyse the SF in different X-ray bands, finding a dependence of the variability on the frequency as $\SF\propto \nu^{-0.15}$, corresponding to a so-called softer when brighter trend. In turn, this dependence allows us to parametrically correct the variability estimated in observer-frame bands to that in the rest frame, resulting in a moderate ($\lesssim 15\%$) shift upwards (V-correction).}
  {Ensemble X-ray variability of AGNs is best described by the structure function. An improper use of the normalised excess variance may lead to an underestimate of the intrinsic variability, so that appropriate corrections to the data or the models must be applied to prevent these effects. }
   
   \keywords{Catalogs - Galaxies: active - Quasars: general - X-rays: galaxies}
   \authorrunning{F.Vagnetti et al.}
\titlerunning{X-ray variability of AGNs}

   \maketitle
%
\section{Introduction}
Variability is a distinctive feature shared by all classes of active galactic nuclei (AGN), occurring in all the wavebands and on different timescales from a fraction of a day up to years. In the X-ray band, variability is observed on  timescales as short as hours, giving insight into the innermost AGN regions, but also on longer timescales, where variability is seen to increase up to at least a few years \citep[see e.g.][]{mark04,vagn11,shem14}.

A large number of studies have investigated the detailed properties of the X-ray variability for many individual AGN, mostly at low redshifts and luminosities \citep[e.g.][]{uttl02,uttl05,pont12}. In cases with sufficient sampling and high signal-to-noise ratios, power spectral density (PSD) analyses have evidenced the typical red-noise character of X-ray variability \citep{gree93,lawr93}.

For AGN in wider intervals of redshift and luminosity, including luminous quasars, faint fluxes and sparse sampling usually prevent detailed individual variability studies, nevertheless, average properties of the X-ray variability have been investigated in several ensemble analyses \citep[e.g.][]{alma00,mann02,paol04,mate07,papa08,vagn11}.

Different methods are used to estimate the variability of these sources and one of the most popular is the normalised excess variance (NXS), which is defined as the difference between the total variance of the light curve and the mean squared error that is normalised for the average of the $N$ flux measurements squared \citep[e.g.][]{nand97,turn99}; see Sect.\,3. This estimator provides an easy way to quantify the AGN variability even for poorly sampled light curves. However, \citet{alle13} have shown that NXS represents a biased estimator of the intrinsic light curve variance, especially when used for individual, sparsely sampled light curves, which results in overestimates or underestimates of the intrinsic variance that depend on the sampling pattern and the PSD slope below the minimum sampled frequency.

Moreover, it has been pointed out that NXS also depends on the length of the monitoring time interval from the red-noise character of the PSD, and decreasing with redshift from the effect of cosmological time dilation \citep[e.g.][] {lawr93,papa08,vagn11}.

The structure function (SF) allows one to compute variability as a function of the rest-frame time lag, and is therefore suitable for ensemble analyses. In \citet[][Paper I]{vagn11}, for example, we used multi-epoch observations of an AGN sample extracted from the XMM-Newton serendipitous source catalogue (XMMSSC) to compute the ensemble X-ray SF. In the present paper, we take advantage of the recent releases of XMMSSC \citep{rose16}, and of the Sloan Digital Sky Survey (SDSS) Quasar Catalogue \citep[][P\^{a}ris et al. in preparation]{pari14}, to compute the normalised excess variance and to update the study of the structure function. Moreover, we show that the latter can be also used to correct the time dilation effect present in the estimates of the former.

The paper is organised as follows. Section 2 describes the data extracted from the archival catalogues. Section 3 computes the light curve duration effect on the NXS estimates. Section 4 updates the SF computation for the new samples. Finally, in Section 5, we discuss and summarise the results.

Throughout the paper, we adopt the cosmology $H_0= 70 {\rm \,km\,s^{-1}\,Mpc^{-1}}$, $\Omega_m=0.3$, and $\Omega_\Lambda=0.7$.

\section{Data}
The XMMSSC catalogue was recently updated to its release 3XMM-DR5 \citep{rose16}, which includes 565\,962 X-ray detections between February 2000 and December 2013, related to 396\,910 unique X-ray sources\footnote{\tt http://xmmssc.irap.omp.eu/}.

\begin{table*}
\centering
\vspace{0.2cm}
\caption{MEXSAS sample$^a$}
\medskip
\begin{tabular}{rclcrrcc}
\hline
$N_{sou}$ &  Name  & $z$  & $\mean{f_X}$ (erg/cm$^2$/s)& $N_{epo}$ & $\Delta t_{rest}$ (days) & NXS & NXScorr \\
\hline
  1     &    3XMM J001716.8-010725 &1.1631  & 1.761E-13  &     3 &101.46 & 0.062  &  0.108    \\
  2     &    3XMM J001731.3-004859 &1.356   & 2.524E-14  &     2 & 93.15 & 0.021  &  0.037    \\
  3     &    3XMM J001808.7-005709 &1.3346  & 3.712E-14  &     3 & 94.01 & 0.131  &  0.231    \\
  4     &    3XMM J004243.0+000201 &1.0822  & 4.150E-14  &     2 & 88.40 & 0.246  &  0.441    \\  
  5     &    3XMM J004316.4+001044 &0.58    & 8.549E-14  &     2 &116.50 & 0.150  &  0.251    \\
  6     &    3XMM J004316.8+001007 &1.5183  & 3.691E-14  &     2 & 73.09 & 0.057  &  0.107    \\ 
\hline
\end{tabular}\\
$^{\rm a}$ The table in its entirety is available at CDS.
\end{table*}

A large number of sources (70\,453) are observed more than once (up to 48 times) for a total of 239\,505 multi-epoch observations, which makes this catalogue very appropriate for variability studies. In Paper I we used the 2XMMi-DR3 release \citep{wats09} that contains 41\,979 multi-epoch sources with a total of 132\,268 observations; thus, with the present release, the number of multi-epoch sources and observations is almost doubled.

To extract a set of X-ray observations for a sample of quasars, we used the software TOPCAT\footnote{\tt http://www.star.bris.ac.uk/$\sim$mbt/topcat/} to cross-correlate the XMMSSC catalogue with the SDSS quasar catalogues, using both Data Release 7 \citep[DR7Q,][]{schn10} and Data Release 12 (DR12Q, P\^aris et al., in preparation). We took into account the quality of the observations, indicated by the parameter {\it SUM\_FLAG}, selecting only  detections with {\it SUM\_FLAG}$<$3, as suggested by the XMMSSC Team. We then searched for coordinate matches within a radius of 5 arcsec, finding 14\,648 matches between the XMMSSC and the SDSS catalogues. Increasing the correlation radius to 10 arcsec produces 15\,095 matches, indicating a possible incompleteness of the order of 3\%. On the other side, repeating the cross-correlation with a set of false coordinates, shifted by 1 arcmin in declination with respect to the true coordinates, we obtained 44 spurious matches, indicating a possible contamination $\sim$0.3\% within the adopted radius.

\begin{figure}
\centering
\resizebox{\hsize}{!}{\includegraphics{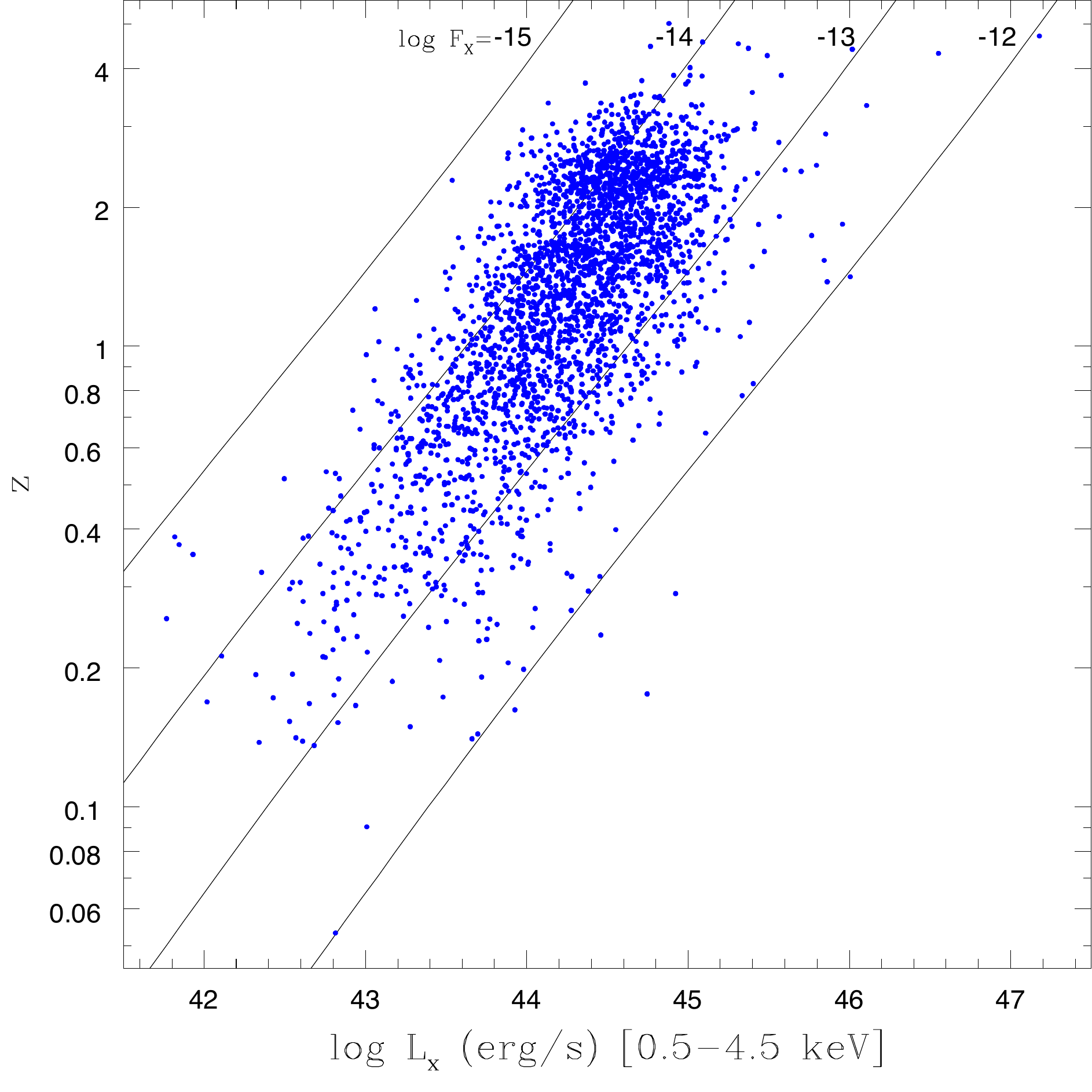}}
\caption{Distribution of the sources in the $L_X$-$z$ plane. The blue dots represent the average values of the X-ray luminosities computed on the available data of each light curve. Lines of constant X-ray flux are also shown.}
\end{figure}

Selecting only sources with multiple matches (at least 2), we found 2112 matches between XMMSSC-DR5 and SDSS-DR7Q, corresponding to 616 unique sources, and 6105 matches between XMMSSC-DR5 and SDSS-DR12Q, corresponding to 2209 unique sources. For 122 sources with 370 X-ray observations, which are found both in DR7Q and DR12Q, we chose the match with the latter, to use a more recent redshift determination. We chose the visual inspection redshift {\it Z\_VI} among the different redshift estimates provided by DR12Q. After an additional check of the parameter {\it SRCID}, which identifies unique sources according to the XMMSSC catalogue, we finally produced a sample of 7837 observations for 2700 sources. To refer  to this sample again in future papers, we will call it Multi-Epoch XMM Serendipitous AGN Sample (MEXSAS). In this work, we use the EP9 band, 0.5-4.5 keV, unless otherwise stated. The main data of the MEXSAS sample are reported in Table 1, where Col. 1 indicates the source serial number $N_{sou}$, Col. 2 the IAU name, Col. 3 the redshift, Col. 4 the average flux in the 0.5-4.5 keV band, in erg/cm$^2$/s, Col. 5 the number of epochs $N_{epo}$ in which the source has been observed, Col. 6 the length of the monitoring time interval in the rest frame, Col. 7 the uncorrected normalised excess variance, and Col. 8 the normalised excess variance corrected after Eq. 9 with $\widehat{\Delta t}=1000$ days and $b=0.12$.

In Fig.1 we show the distribution of the sources in the luminosity-redshift plane, where $L_X$ indicates the luminosity in the X-ray band 0.5-4.5 keV, which is computed from the corresponding flux in the EP9 band and directly extracted from the XMMSSC catalogue, by adopting a photon index $\Gamma=1.7$.

It is to be remarked that the EP9 flux errors available in the previous release XMMSSC-DR4 were wrongly estimated\footnote{{\tt http://xmmssc-www.star.le.ac.uk/Catalogue/\\ xcat\_public\_3XMM-DR4.html\#watchouts.} This watchout was indeed published after the XMMSSC Team had been alerted of the problem by our group.}. This problem was not present in DR3 release and has been corrected in DR5 release, as shown in Fig.2.

\begin{figure}
\centering
\resizebox{\hsize}{!}{\includegraphics{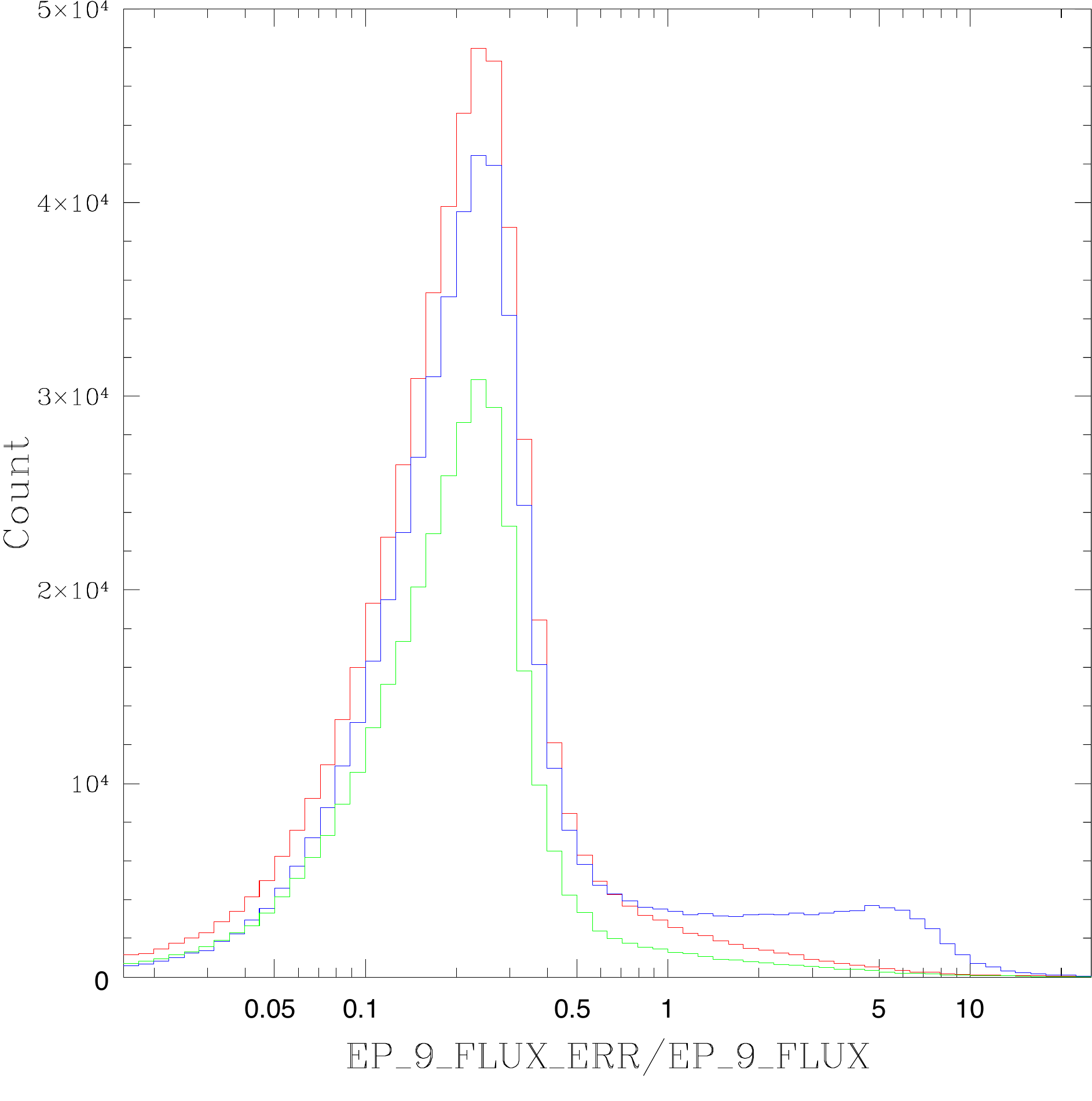}}
\caption{Histograms of the relative EP9 flux errors in the XMMSSC releases. DR3, green; DR4, blue; DR5, red. Anomalously large errors are present for a subset of the DR4 catalogue.}
\end{figure}

\section{The excess variance and the light curve duration effect}
The normalised excess variance is defined by the equation
\begin{equation}
\sigma^2_{NXS}=\frac{S^2-\sigma_{n}^2}{\mean{f}^2}~,
\end{equation}

\noindent
where $\mean{f}=\sum_{i=1}^N f_i/N$ is the mean flux computed over the available flux measures $f_i$ of the same source, $S^2=\frac{1}{N-1}\sum_{i=1}^N f_i^2-\mean{f}^2$ is the total variance of the light curve, while $\sigma_n^2=\sum_{i=1}^N \sigma_i^2/N$ is the mean square photometric error associated with the measured fluxes $f_i$. 

\begin{figure}
\centering
\resizebox{\hsize}{!}{\includegraphics{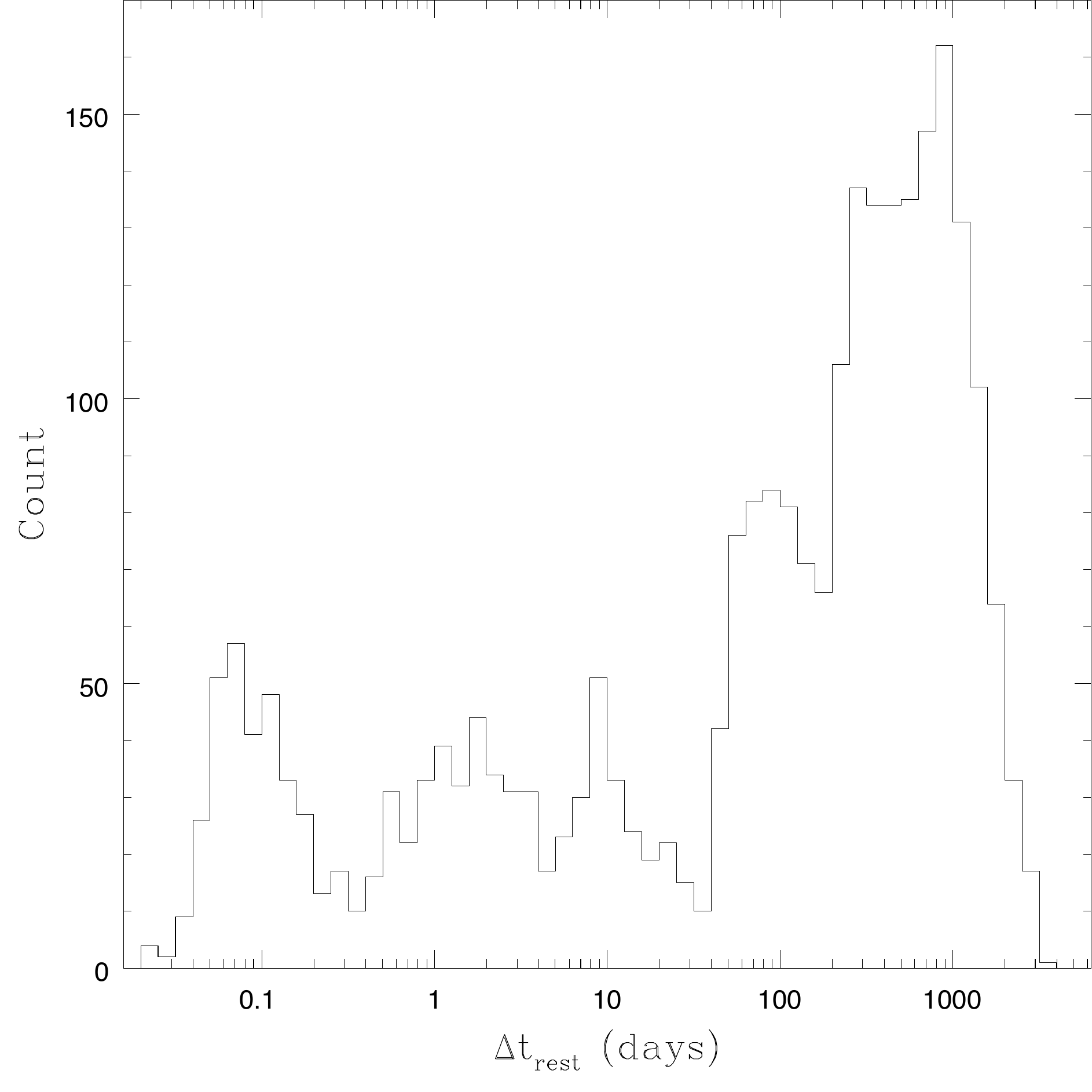}}
\caption{Distribution of the light curve durations in the rest frame, $\Delta t_{rest}$, for the MEXSAS sample.}
\end{figure}

Because NXS is an estimate of the average variability within the monitoring time interval $\Delta t_{obs}$ provided by the light curve, and because variability increases with the rest-frame time lag $\tau$ \citep[e.g.][]{mark04,vagn11}, we expect that NXS also increases with the length of the monitoring time in the rest frame of the source, $\Delta t_{rest}=\Delta t_{obs}/(1+z)$, whose distribution is shown in Fig. 3.

We then compute $\sigma^2_{NXS}$ for the EP9 fluxes of all the 2700 sources of the MEXSAS sample, and report them in Fig. 4, as a function of the number of epochs $N_{epo}$ sampled by the light curve. We notice two points: first, the large dispersion of the NXS values for poorly sampled light curves that quickly decreases for increasing $N_{epo}$, and, second, the presence of negative values that are also more frequent for small $N_{epo}$. In fact, NXS is computed with respect to the light curve average flux $\mean{f}$, which differs from the intrinsic mean $\mu$ \citep[see, e.g.][]{alle13}, and its expected deviation is larger for smaller numbers of sampled data. Moreover, the observed variance can be smaller than the error, resulting in a negative NXS that is more probable when the mean is less well estimated, so again for small $N_{epo}$. 

\begin{figure}
\centering
\resizebox{\hsize}{!}{\includegraphics{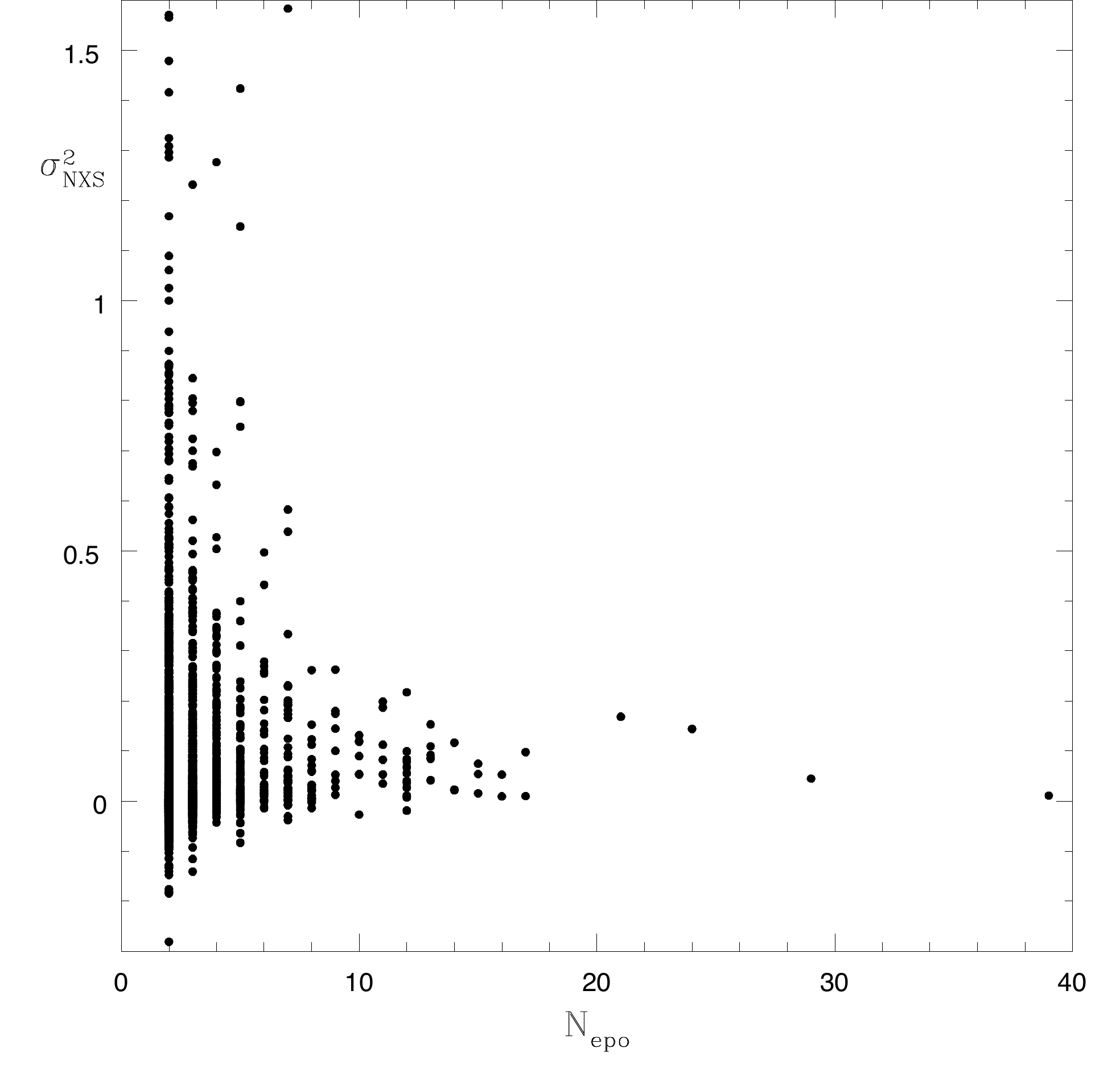}}
\caption{NXS as a function of the number of epochs in the light curve.}
\end{figure}

\begin{figure}
\centering
\resizebox{\hsize}{!}{\includegraphics{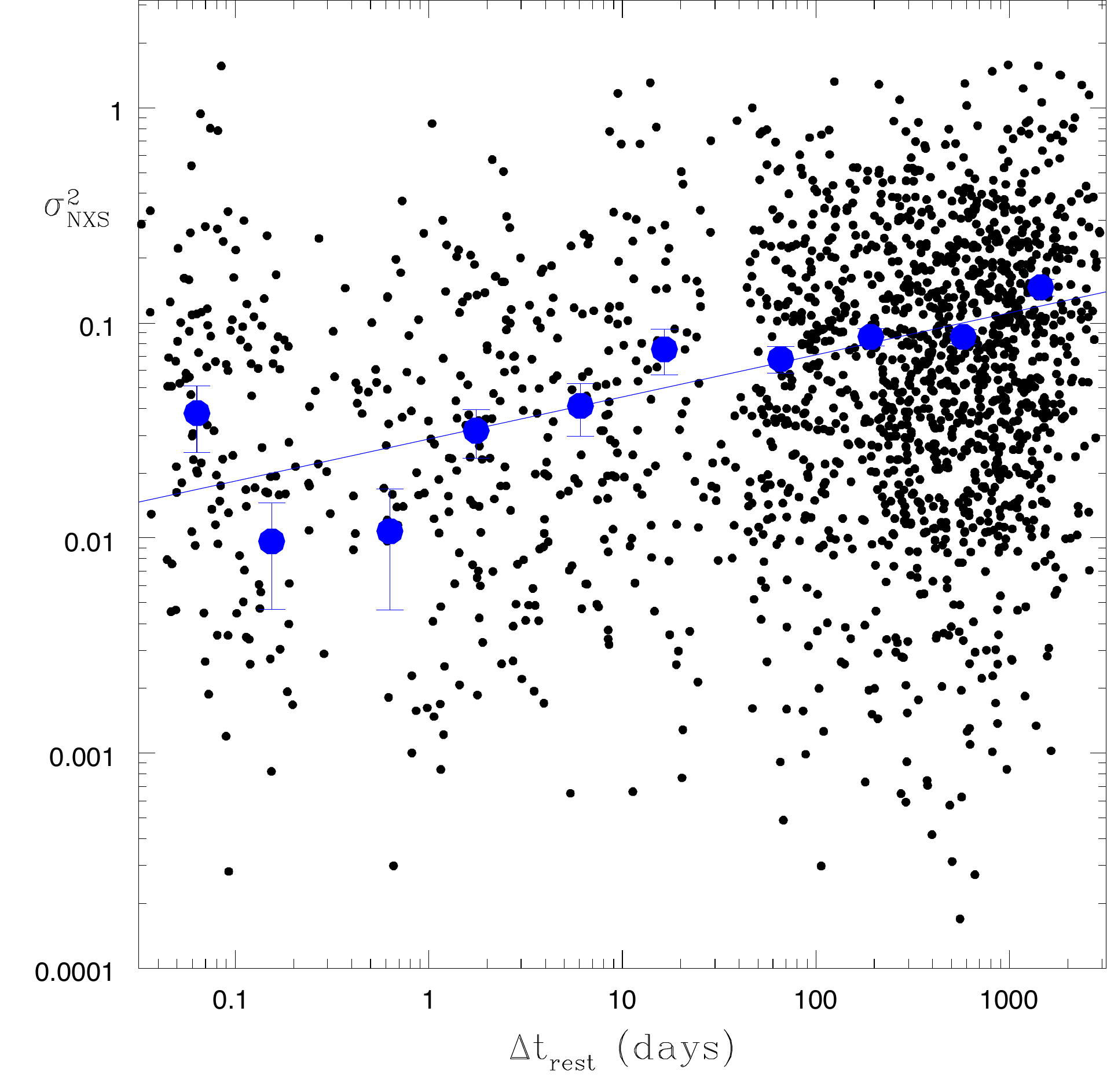}}
\caption{NXS as a function of the light curve duration in the rest frame, $\Delta t_{rest}$. Black dots represent individual NXS values. Blue circles are ensemble averages in bins of $\Delta t_{rest}$.
The solid blue line shows a linear least-squares fit to the logarithms of the binned values with slope $a=0.196\pm0.040$. The Pearson correlation coefficient is $r=0.84$, with null probability $P(>|r|)=0.008$.}
\end{figure}

We now show in Fig. 5 the log of the excess variance as a function of the log of the rest-frame duration. Following \citet{alle13}, ensemble estimates of NXS are to be preferred to the individual values. We report ensemble averages in bins of $\Delta t_{rest}$, also including negative contributions. In fact, the removal of negative values might skew the distribution if not equally spread over the whole population. Individual values of NXS are also shown, when NXS$>0$. A clear increase appears, as expected. The binned values can be fitted by a straight line with slope $a=0.196\pm0.040$, i.e. a power-law $\sigma^2_{NXS}\propto\Delta t_{rest}^{0.20}$. The Pearson correlation coefficient is $r=0.84$ with null probability $P(>|r|)=0.008$.

We notice some possible sampling effects. First, different light curves are sampled with different patterns so this can introduce systematic differences, although this effect is not larger than 50\% for PSD slopes that are shallower than $-2$ \citep{alle13}. Second, when sampling long timescales, the red noise leak is smaller than for short timescales, and this might make the intrinsic slope of our fitted line steeper. However, our aim here is to show that NXS increases with the light curve duration $\Delta t_{rest}$. The precise value of the slope might be improved taking these additional effects into account.

\section{ Structure function}
The structure function works in the time domain and is very helpful for an ensemble analysis of the variability, even for poor sampling of the individual sources, as in the present case. It is often used in the optical band \citep[e.g.][]{trev94,vand04,wilh08,baue09,macl12} and is used less often in the X-rays, where the only ensemble analysis was performed by us in Paper I, in which we defined
\begin{equation}
\SF(\tau)\equiv\sqrt{\frac{\pi}{2}\langle|\log f_X(t+\tau)-\log f_X(t)|\rangle^2-\sigma^2_{noise}}\quad ,
\end{equation}

\noindent
$f_X(t)$ and $f_X(t+\tau)$ as two measures of the flux, in a given X-ray band, at two epochs differing by time lag $\tau$ in the rest frame. The term $\sigma_{noise}^2=\mean{\sigma_n^2(t)+\sigma_n^2(t+\tau)}$ is the quadratic contribution of the photometric noise to the observed variations \citep[see also the discussion by][]{kozo16}. The average is computed within an appropriate bin of time lag around $\tau$. The average of the absolute value of the variations was adopted because it is less sensitive to outliers and, in analogy with the expression introduced by \citet{di-c96}, in the optical. In the following, however, we also use  the other standard expression first introduced by \citet{simo85} 
\begin{equation}
\SF(\tau)\equiv\sqrt{\langle[\log f_X(t+\tau)-\log f_X(t)]^2\rangle-\sigma^2_{noise}}\quad .
\end{equation}

\noindent
The two expressions are equivalent if the variations follow a Gaussian distribution and  the number of measured variations is large enough. If one or both the conditions are not fulfilled, the expression of Eq.3 is sometimes preferred because it is directly related to other statistical quantities such as the autocorrelation function and the variance, although the differences are relatively small \citep[see e.g.][]{baue09}.

\subsection{Updated ensemble SF}
\begin{figure}
\centering
\resizebox{\hsize}{!}{\includegraphics{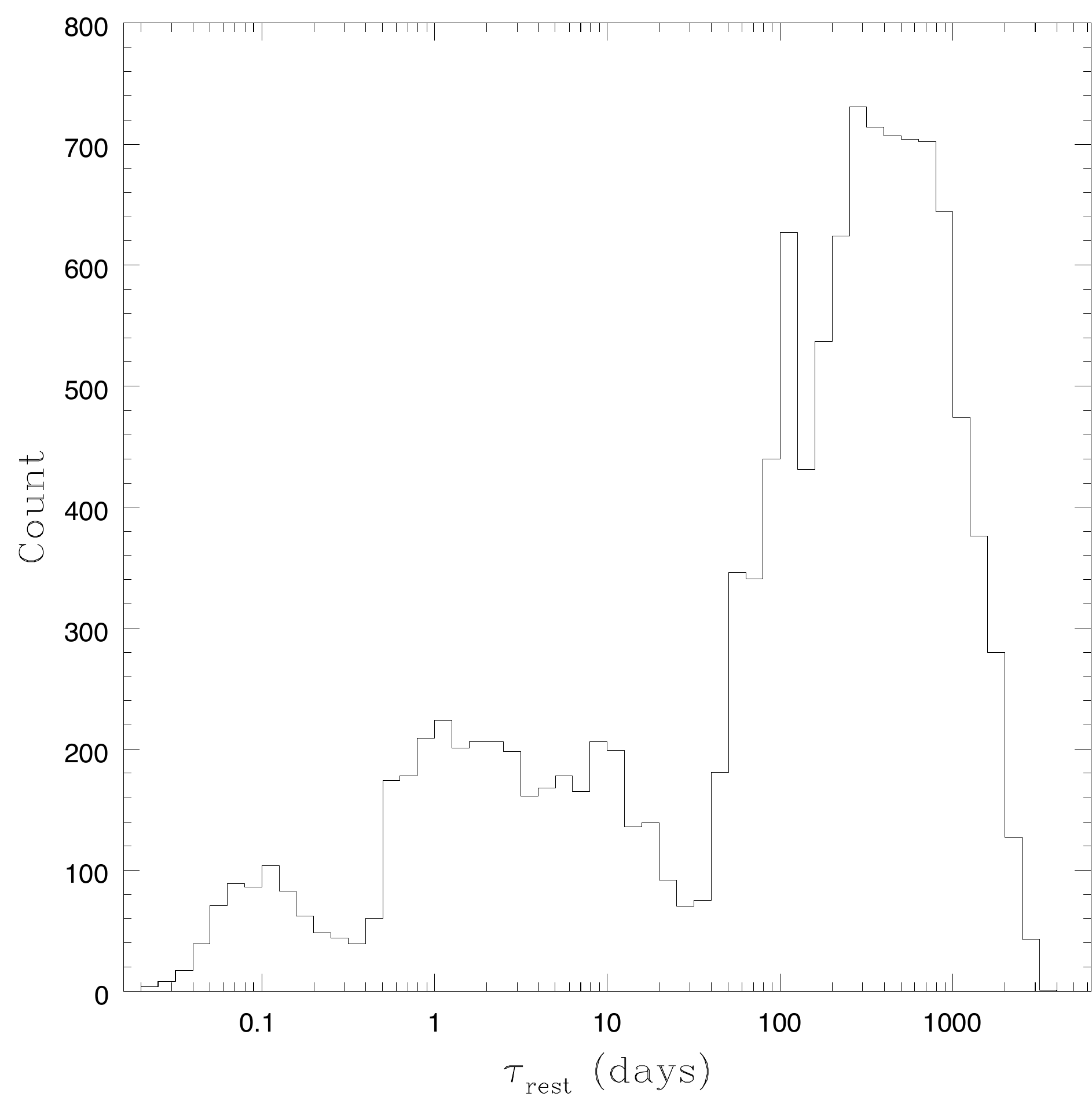}}
\caption{Histogram of the rest-frame lag times for the MEXSAS sample.}
\end{figure}
We show in Fig. 6 the distribution of the rest-frame lag times $\tau_{rest}$ for the flux variations contributing to the computation of the SF for the MEXSAS sample. 
The histogram looks similar to that shown in Fig.  3 for the light curve durations, but it is much more populated due to all the possible combinations of pairs of observations.

We computed the SF for the MEXSAS sample, again using the EP9 fluxes. The result is shown in Fig. 7, using both Eq. 2 (red symbols and lines) and Eq. 3 (blue symbols and lines), with the SF computed in bins of $\log\tau_{rest}$. The representative points of the bins are centred weighting the individual lag values falling in each bin to take account of the non-uniform distribution of $\tau_{rest}$ shown in Fig. 6. The SF has been fitted by a power-law $\SF=k\tau^b$, through a linear least-squares fit of the logarithms, weighted with the number of individual lag values falling in each bin.
\begin{figure}
\centering
\resizebox{\hsize}{!}{\includegraphics{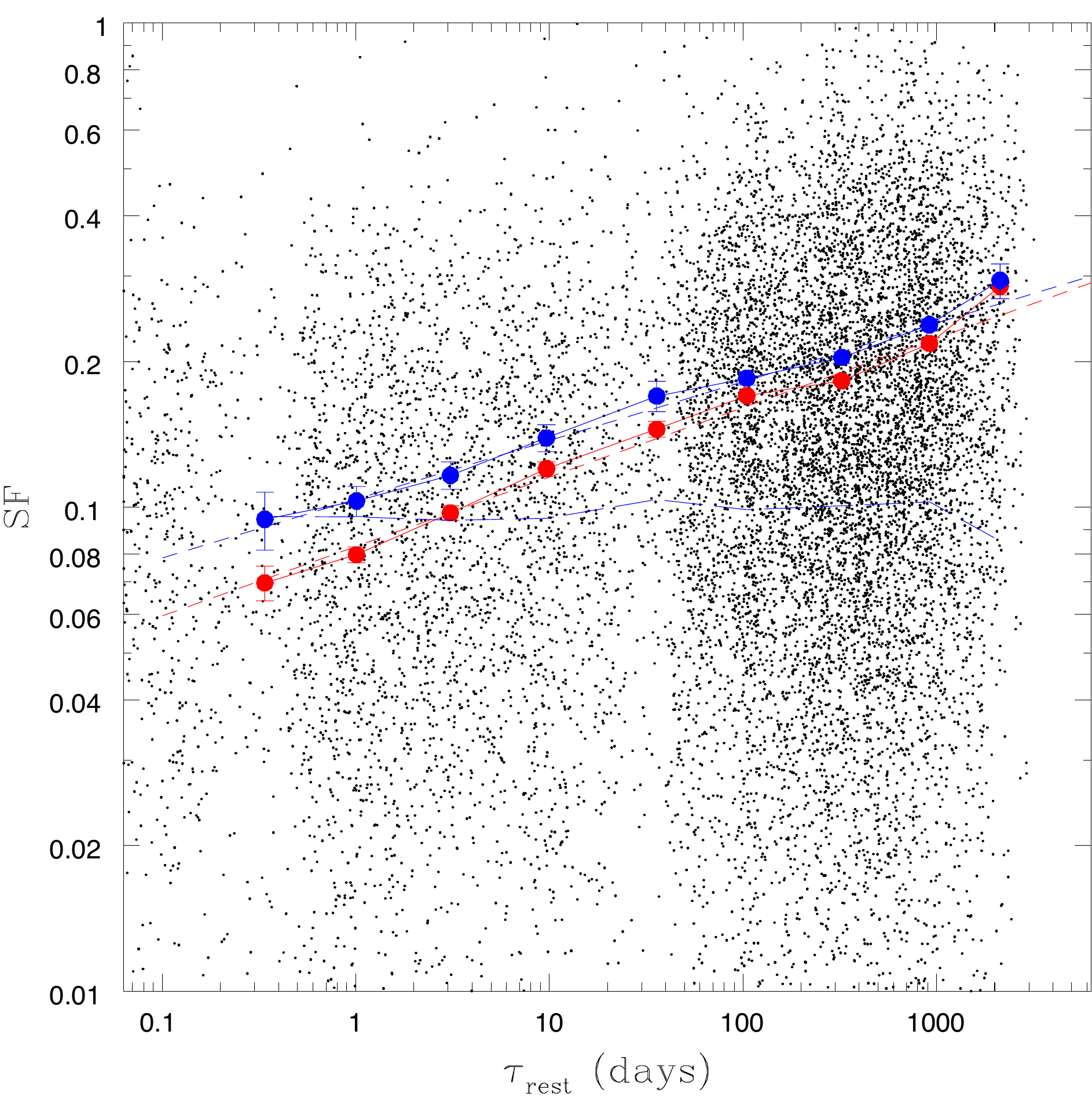}}
\caption{Structure function for the MEXSAS sample. Red points and continuous lines represent the SF computed for the EP9 flux variations, according to Eq. 2; blue points and continuous lines refer to the EP9 band, using Eq. 3. Red and blue short-dashed lines indicate the corresponding least-squares fits. Blue long-dashed line indicates the contribution of the photometric errors (the same for Eq. 2 and 3). Black dots represent the variations for the individual pairs of measurements contributing to the SF.}
\end{figure}

The SF computed with the average of the square differences appears slightly flatter ($b=0.121\pm0.004$) than the SF obtained using the average of the absolute values ($b=0.143\pm0.006$), suggesting that the two expressions are not equivalent. In fact, we checked the distributions of our variations of $\log f_X$ for normality, applying a Kolmogorov-Smirnov test in each of the bins used in Fig.7, always finding small probabilities that range from a few percent to $10^{-11}$ depending on the bin population. Thus our distributions are not Gaussian and the expression of Eq.3 is preferred.

Including normalisation, the SF computed with Eq. 3 is given by $\log\SF=(0.121\pm0.004)\log\tau_{rest}-(0.983\pm0.010)$, so that  its value at 1000 days is $\approx 0.24$.

This updates the previous ensemble SF of Fig. 5 of Paper I, which was derived from a much smaller sample.

\subsection{Correction of the NXS}
We now want to use the dependence of the variability on the time lag, expressed through the SF, to estimate the expected value of the NXS in a given monitoring interval $\Delta t_{obs}$.
We first rewrite Eq. 3, neglecting the photometric error
\begin{equation}
\SF(\tau)=\sqrt{\langle[\log f_{int}(t+\tau)-\log f_{int}(t)]^2\rangle}=\sqrt{\langle(\delta\log f_{int})^2\rangle}\,,
\end{equation}
\noindent
meaning that we refer to the intrinsic variations $\delta\log f_{int}$.
Similarly, we rewrite Eq.1, also neglecting the photometric error with the same meaning as above, as follows:
\begin{equation}
\sigma^2_{NXS}=\frac{S^2}{\mean{f_{int}}^2}=\frac{\mean{\delta f_{int}^2}}{\mean{f_{int}}^2}\approx 
\frac{\mean{(\delta\log f_{int})^2}}{(\log e)^2}\,.
\end{equation}
\noindent
Here the average must be computed within the monitoring time interval $\Delta t_{obs}$. Both Eqs. 4 and 5 are expressed in terms of average square variations of $\log f_{int}$, thus we can rewrite
\begin{equation}
\sigma^2_{NXS}\approx\frac{\mean{S\negthinspace F^2}}{2(\log e)^2}\,,
\end{equation}
\noindent
where the factor 1/2 accounts for the two independent measures contributing to each SF flux difference, and the average must be computed within the rest-frame time interval $\Delta t_{rest}=\Delta t_{obs}/(1+z)$.
\begin{figure}
\centering
\resizebox{\hsize}{!}{\includegraphics{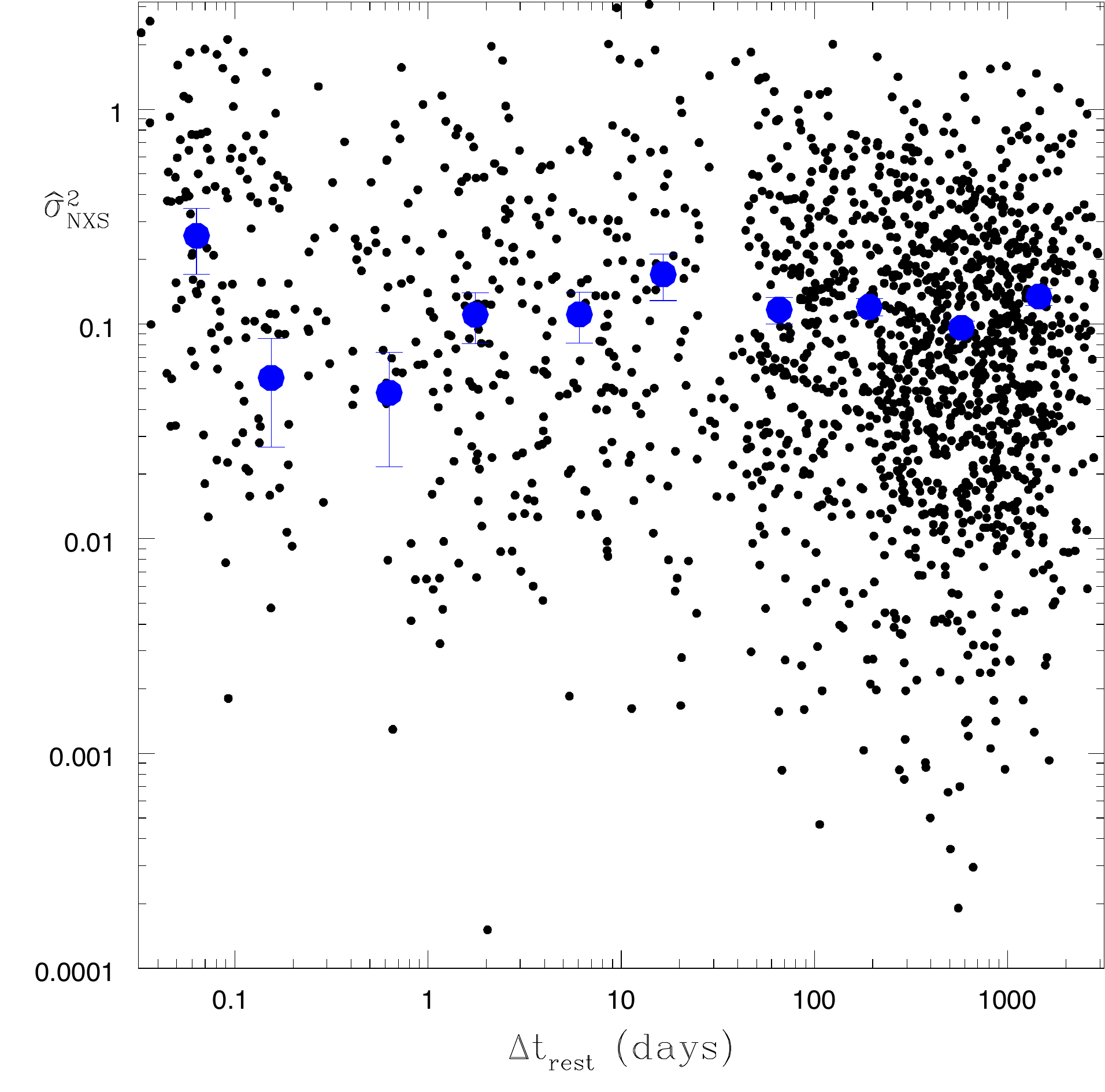}}
\caption{Corrected NXS as a function of the light curve duration in the rest frame, $\Delta t_{rest}$, using $b=0.10$ and $\widehat{\Delta t}=1000$ days for individual values (black dots) and for binned averages (blue cicles). There is no correlation with $\Delta t_{rest}$.}
\end{figure}

Adopting now a functional form of the SF, for example a power-law $S\negthinspace F=k\tau^b$ as in Paper I, we compute the average as follows:
\begin{equation}
\mean{\SF^2}=\frac{1}{\Delta t_{rest}}\int_0^{\Delta t_{rest}}k^2\tau^{2b}d\tau=\frac{[\SF(\Delta t_{rest})]^2}{2b+1}\,,
\end{equation}
\noindent
and finally we obtain
\begin{equation}
\sigma^2_{NXS}=\frac{k^2\Delta t_{rest}^{2b}}{2(2b+1)(\log e)^2}=\frac{k^2}{2(2b+1)(\log e)^2}\left(\frac{\Delta t_{obs}}{1+z}\right)^{2b}\,,
\end{equation}
\noindent
which shows that NXS is also expected to increase with a power law of the monitoring time interval. If this is expressed in the observer frame, an obvious dependence on the redshift is also found.
Using Eq. 8, it is now possible to correct for the duration effect, extrapolating the measured NXS values to a fixed rest-frame time interval $\widehat{\Delta t}$ as follows:
\begin{equation}
\widehat{\sigma}^2_{NXS}=\sigma^2_{NXS}\left({\widehat{\Delta t}/\Delta t_{rest}}\right)^{2b}\,.
\end{equation}
\noindent
This correction can be applied to a given set of NXS values to obtain new estimates referred to a uniform duration, adopting a previously determined SF exponent from literature, for example $b=0.10$ from Paper I. Choosing $\widehat{\Delta t}$=1000 days, the corrected values of $\widehat{\sigma}^2_{NXS}$ are shown in Fig. 8. There is no correlation with $\Delta t_{rest}$, the Pearson correlation coefficient is $r=0.12$ and the probability of obtaining this by chance is $P(>|r|)=0.70$. The choice of the value $b=0.12$ from the updated SF of the present paper would give similar results [$r=-0.16, P(>|r|)=0.60$].

On the other hand, the possible change in slope of the PSD would affect this relation for the shortest timescales; however the break is usually $<100$ days (for black hole masses $M_{BH}<10^9$; \citet{gonz12}), so this effect is only relevant for the most massive BHs.

\subsection{Dependence on X-ray luminosity and redshift}
\begin{figure}
\centering
\resizebox{\hsize}{!}{\includegraphics{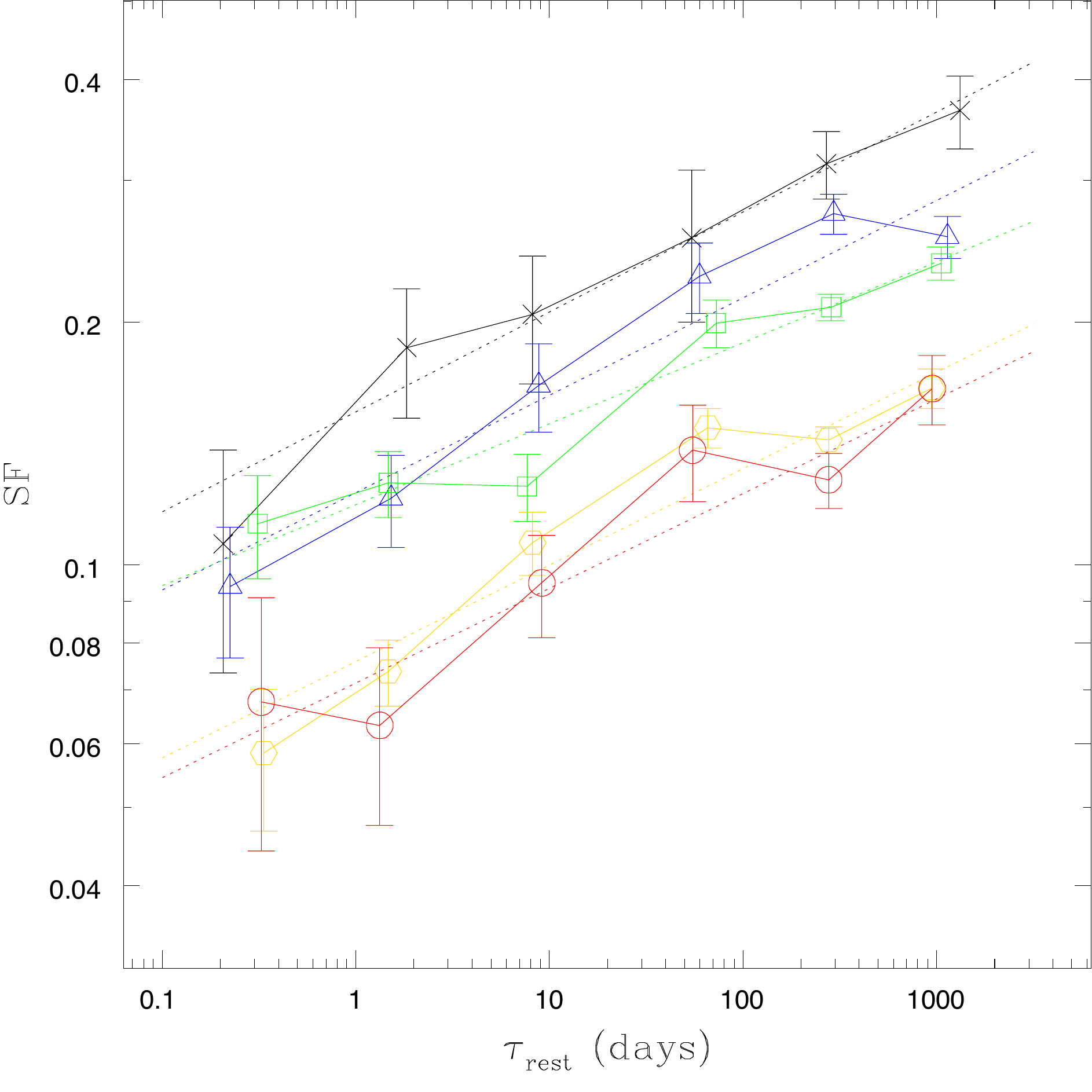}}
\caption{Structure function in bins of X-ray luminosity. Black lines and crosses denote $10^{43}{\rm\,erg/s}<L_X\leq 10^{43.5}{\rm\,erg/s}$; blue lines and triangles denote $10^{43.5}{\rm\,erg/s}<L_X\leq 10^{44}{\rm\,erg/s}$; green lines and squares indicate $10^{44}{\rm\,erg/s}<L_X\leq 10^{44.5}{\rm\,erg/s}$; yellow lines and hexagons represent $10^{44.5}{\rm\,erg/s}<L_X\leq 10^{45}{\rm\,erg/s}$; and red lines and circles indicate $10^{45}{\rm\,erg/s}<L_X\leq 10^{45.5}{\rm\,erg/s}$.}
\end{figure}
We then update the analysis of the SF as a function of the X-ray luminosity in a  similar way as that performed in Paper I, dividing our sample in luminosity bins. Our present sample is much richer compared to that used in Paper I and allows us to extend our analysis to lower luminosities to between $L_X=10^{43}$ erg/s and $L_X=10^{45.5}$ erg/s. At variance with Paper I, for the present sample we find (see Fig. 9) almost uniform slopes of the SF in the different luminosity bins, while the normalisation strongly depends on $L_X$. This work differs from Paper I, where we found slopes changing with $L_X$, in that we have a much richer sample of 2700 sources compared to 412 in the fist paper. In that case, the number of unbinned SF points contributing to the shortest time-lag bin was small, and therefore only a few points contributed, once they were further divided in bins of luminosity; this resulted in a large dispersion of mean SF values in bins of luminosity, thereby artificially producing a dispersion in the slopes.
\begin{figure}
\centering
\resizebox{\hsize}{!}{\includegraphics{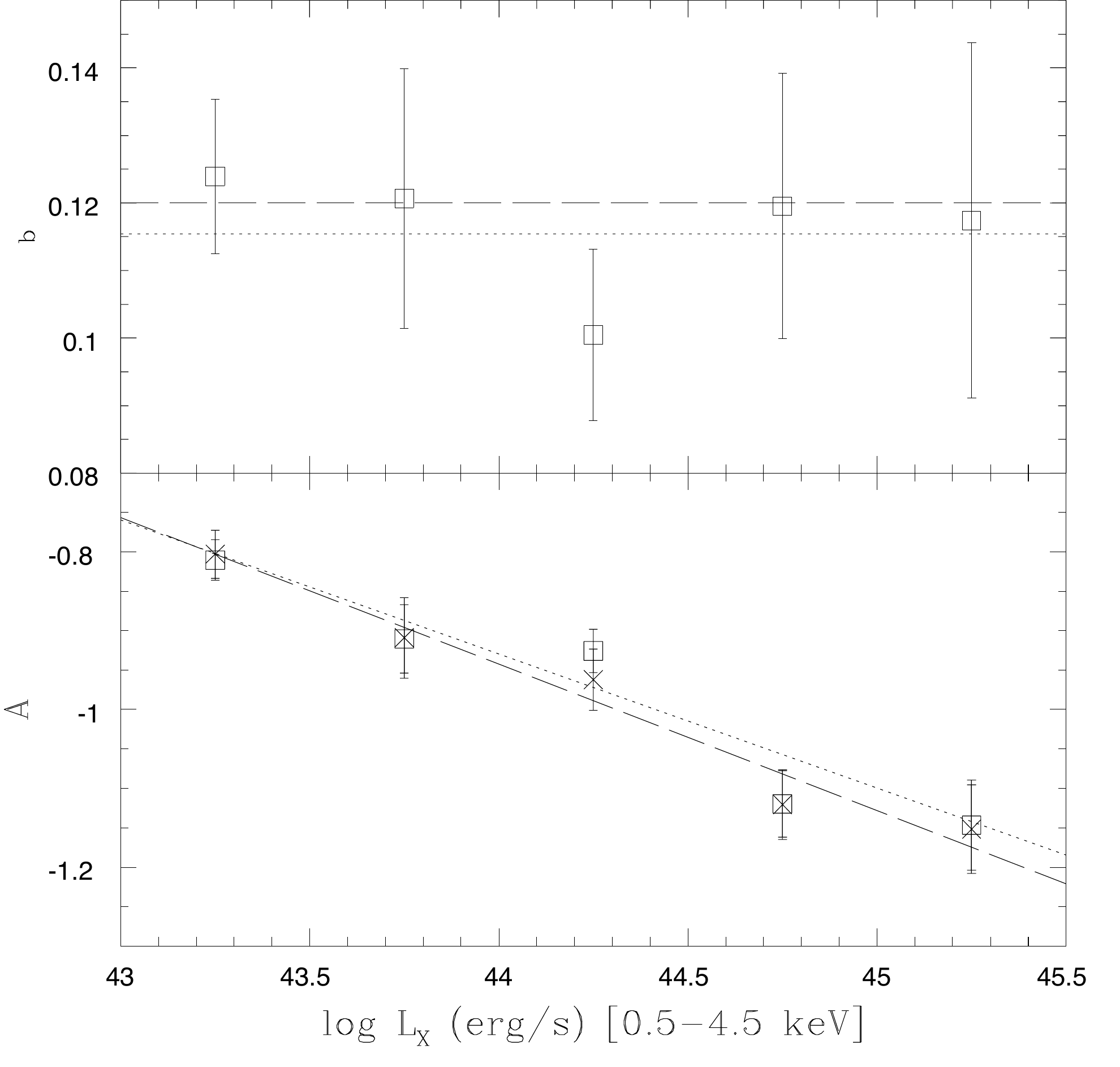}}
\caption{Structure function parameters as functions of the X-ray luminosity. {\it Upper panel}: slope $b$; the dotted line indicates the average $\mean{b}=0.115$; the dashed line indicates the fixed value $b=0.12$, adopting the same dependence as in the general SF of Fig. 7. {\it Lower panel}: the intercept $A$. The open squares represent the values derived by the SFs of Fig. 9, with free $A$ and $b$. The crosses are the values derived with fixed $b=0.12$. The corresponding fits are shown with dotted lines (free $b$, $A=(6.55\pm1.42)-(0.17\pm0.03)\log L_X$) and dashed lines (fixed $b$, $A=(7.24\pm0.81)-(0.19\pm0.02)\log L_X$).}
\end{figure}

Describing the SF as $\log\SF=A+b\log\tau_{rest}$, we show in Fig. 10 the values of the slopes $b$ and the intercepts $A$ for the different luminosity bins. The slopes are almost constant with an average value $\mean{b}=0.115$, and are compatible within $2\sigma$ with the slope $b=0.12$ of the overall sample shown in Fig. 7. The intercepts are clearly anti-correlated with $L_X$ (correlation coefficient $r=-0.96$), and a weighted least-squares fit gives $A=(6.55\pm1.42)-(0.170\pm0.032)\log L_X$. Assuming a fixed slope, $b=0.12$, changes the estimates of the intercepts with a fit $A=(7.24\pm0.81)-(0.186\pm0.018)\log L_X$. This corresponds to values of the structure function at 1000 days decreasing approximately from 0.35 to 0.15 for increasing $L_X$ within the adopted bins.

\begin{figure}
\centering
\resizebox{\hsize}{!}{\includegraphics{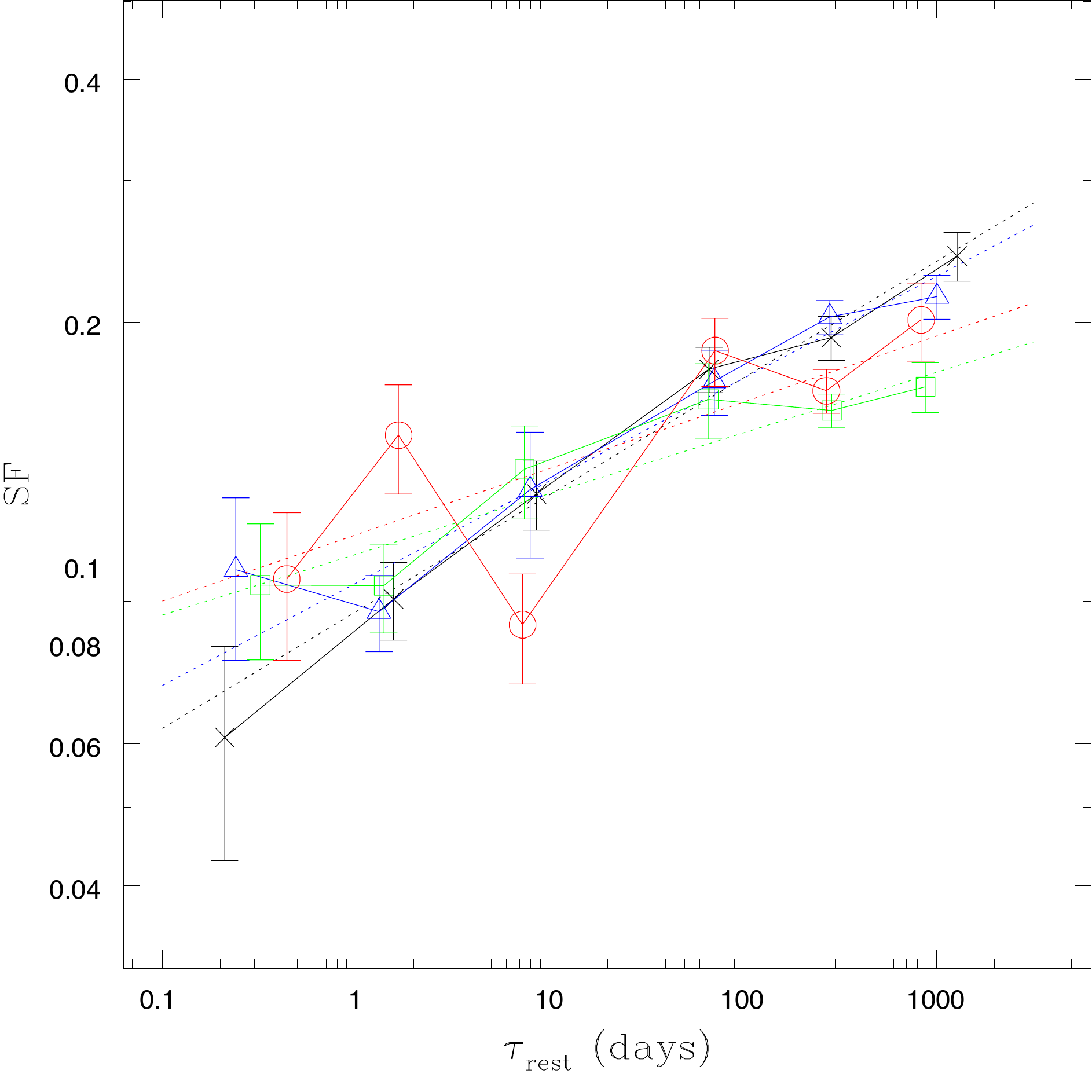}}
\caption{Structure function in bins of redshift. Black lines and crosses represent $0<z\leq 1.15$; blue lines and triangles indicate $1.15<z\leq 1.7$; green lines and squares denote $1.7<z\leq 2.3$; and red lines and circles indicate $2.3<z\leq 3.4$.}
\end{figure}

Fig. 11 shows the SF divided in bins of redshift. We only considered sources with $10^{44}{\rm\,erg/s}<L_X\leq 10^{45}{\rm\,erg/s}$  to reduce the observational correlation between redshift and luminosity (see Fig. 1). Four bins of redshift are considered: $0<z\leq 1.15$, $1.15<z\leq 1.7$, $1.7<z\leq 2.3$, and $2.3<z\leq 3.4$. The SFs are largely overlapped with no evidence of a change in normalisation. A weak flattening of the slopes for higher redshifts might be suggested. However, at variance with Paper I, where we found a significant partial correlation coefficient of variability with redshift (compensating for the change in $L_X$), we now obtain $r_{Vz,L}=0.05$, which we interpret as no evidence of a dependence on redshift.

In addition, we note that $z$ dependence could be affected by the different rest-frame energy ranges probed at different redshifts. This is further discussed in Sect. 4.4.1.

\subsection{Dependence on the emission band and spectral variability}
Variability can of course also depend on the emission band. Results for individual Seyfert galaxies typically show a decrease of variability towards harder X-ray bands \citep[e.g.][]{sobo09}, corresponding to a softer when brighter spectral variability. The same trend might also hold for quasars and high luminosity AGNs; for example \citet{gibs12} find a softer when brighter behaviour for a small sample of 16 radio-quiet, non-BAL quasars extracted from the Chandra public archive. For our sample, we can investigate an ensemble behaviour indirectly, computing the structure function in different X-ray bands, while a more direct analysis of the photon index variations will be presented in a future paper (Serafinelli et al. in preparation).

\begin{figure}
\centering
\resizebox{\hsize}{!}{\includegraphics{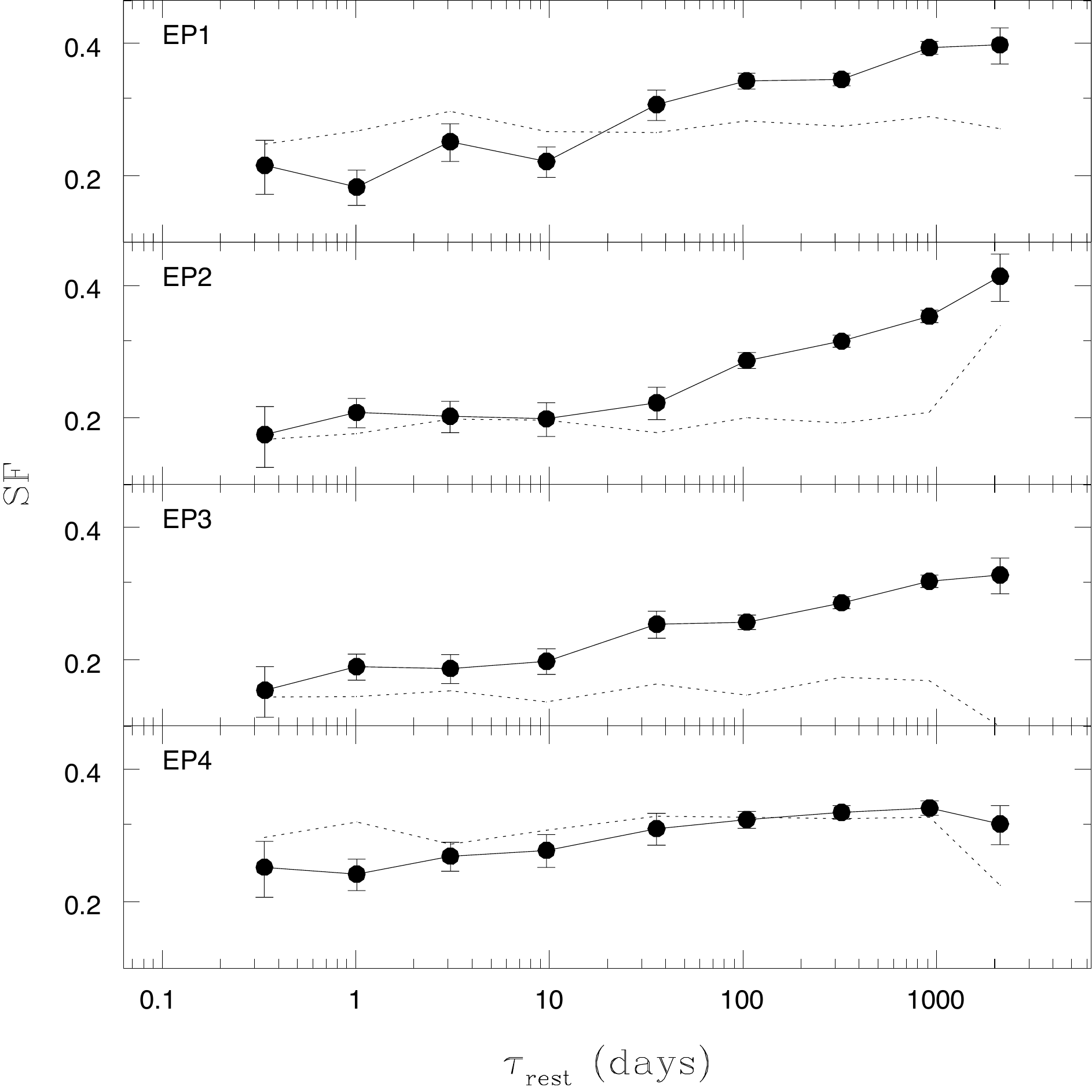}}
\caption{Structure function for the XMM-Newton bands EP1, EP2, EP3, and EP4 (filled circles and continuous lines). Also shown is the contribution of photometric errors, which has been subtracted from the observed variations according to Eq. 3 (dotted lines).}
\end{figure}
We show in Fig. 12 the structure functions for the MEXSAS sample for each of the XMM-Newton spectral bands 0.2-0.5 keV (EP1), 0.5-1 keV (EP2), 1-2 keV (EP3), and 2-4.5 keV (EP4). We do not show the 4.5-12 keV band (EP5), which is strongly affected by photometric errors and is less reliable. The figure shows the structure functions computed after Eq. 3 (filled symbols and continuous lines) and the contribution of the photometric errors (dotted lines), which has been subtracted accordingly. The contribution of the errors is relatively high compared to the wider EP9 band (see Fig. 7) because of the smaller photon counts in these narrower bands, and thus these structure functions are more reliable for lags larger than $\sim 30$ days. Furthermore, we notice that there is a regular trend of decreasing variability from EP1 to EP3, while there is a more complex behaviour for band EP4 with some increase and flattening. Considering only the bands EP1, EP2, EP3, and averaging the SF in the lag interval 100 days $\leq\tau_{rest}\leq$ 1000 days, we find a dependence on the emission frequency given by
\begin{equation}
\log\SF=(2.06\pm 0.15)-(0.15\pm0.01)\log\nu\ .
\end{equation}

In turn, the dependence of variability on the emission frequency can be connected to the spectral variations, as was carried out by \citet{trev02} for the optical band through the definition of the spectral variability parameter
\begin{equation}
\beta=\frac{\Delta\alpha}{\Delta\log f_\nu}\,,
\end{equation}
\noindent
which relates the temporal changes of the spectral index\footnote{Defined after $F_\nu\propto\nu^\alpha$} with those of the monochromatic flux. Values $\beta>0$ correspond to a harder when brighter behaviour, typically observed in the optical band. In the X-ray band, the $\beta$ parameter can be rewritten as
\begin{equation}
\beta=-\frac{\Delta\Gamma}{\Delta\log f_X}
\end{equation}
\noindent
in terms of changes of the photon index $\Gamma$\footnote{$P(E)\propto E^{-\Gamma}$, $\Gamma=1-\alpha$} and of the corresponding flux in the considered X-ray band, $f_X$. Negative $\beta$ values are expected for a softer when brighter behaviour.

Consider now a logarithmic flux variation in a given X-ray band, $\Delta\log f_X$, which is essentially the structure function $\SF$. When the photon index changes by $\Delta\Gamma$, variations at different frequencies separated by $\delta\log\nu$ change as
$$\delta\,\SF=-\Delta\Gamma\cdot\delta\log\nu\,.$$

\noindent
From the definition, Eq. 12, we have
$$\Delta\Gamma=-\beta\cdot\Delta\log f_X\approx-\beta\cdot\SF\ .$$

\noindent
Thus we have $\delta\,\SF=\beta\cdot\SF\,\delta\log\nu$ and 

$$\frac{\delta\log\SF}{\delta\log\nu}\simeq\frac{1}{\delta\log\nu}\log e\cdot\frac{\delta\,\SF}{\SF}=\beta\log e\ ,$$

\noindent
so that from Eq. 10 we can estimate 
\begin{equation}
\beta\simeq 2.3\cdot\delta\log\SF/\delta\log\nu\approx -0.35\pm0.02\ .
\end{equation}

\noindent
This value is also in approximate agreement with a direct analysis of the photon index variations that is in progress (Serafinelli et al., in preparation).

\subsubsection{V-correction.}
The dependence of variability on frequency also implies that variability in the rest frame is not the same as estimated in the observer frame. Our analysis of the variability is based on data tabulated in observer-frame bands. We cannot fix the rest-frame band for AGNs at different redshifts. However, we can use the estimated spectral variability, Eq. 13, to simulate the shift from observer frame to rest frame, as follows.

For a source at redshift $z$, we are measuring variability in a rest-frame band shifted by $\delta\log\nu=\log(1+z)$, so that 
$$\delta\,\SF=\beta\cdot\SF\cdot\log(1+z)\,.$$

\noindent
We can derive
$$\delta\log\SF\simeq\log e\cdot\frac{\delta\,\SF}{\SF}\simeq\log e\cdot\beta\log(1+z)\,.$$

\noindent
The average effect for a sample is a downwards shift (for $\beta<0$), so that to correct the SF we should apply the opposite upwards shift, which we call V-correction:
\begin{equation}
\textrm{V-corr}\equiv-\mean{\delta\log\SF}\simeq -\log e\cdot\beta\mean{\log(1+z)}\ .
\end{equation}

\noindent
We note here that the standard K-correction has no effect on our SFs because fluxes before and after a variation are affected by the same $z$-dependent factor for a given source, so that the corresponding logarithmic change is not altered.

Taking $\beta$ from Eq. 13 and $\mean{\log(1+z)}\simeq 0.38$, we estimate for our sample $\textrm{V-corr}\simeq 0.06$ for the logarithm of the SF, or a $\lesssim 15\%$ correction to the SF itself.

\begin{figure}
\centering
\resizebox{\hsize}{!}{\includegraphics{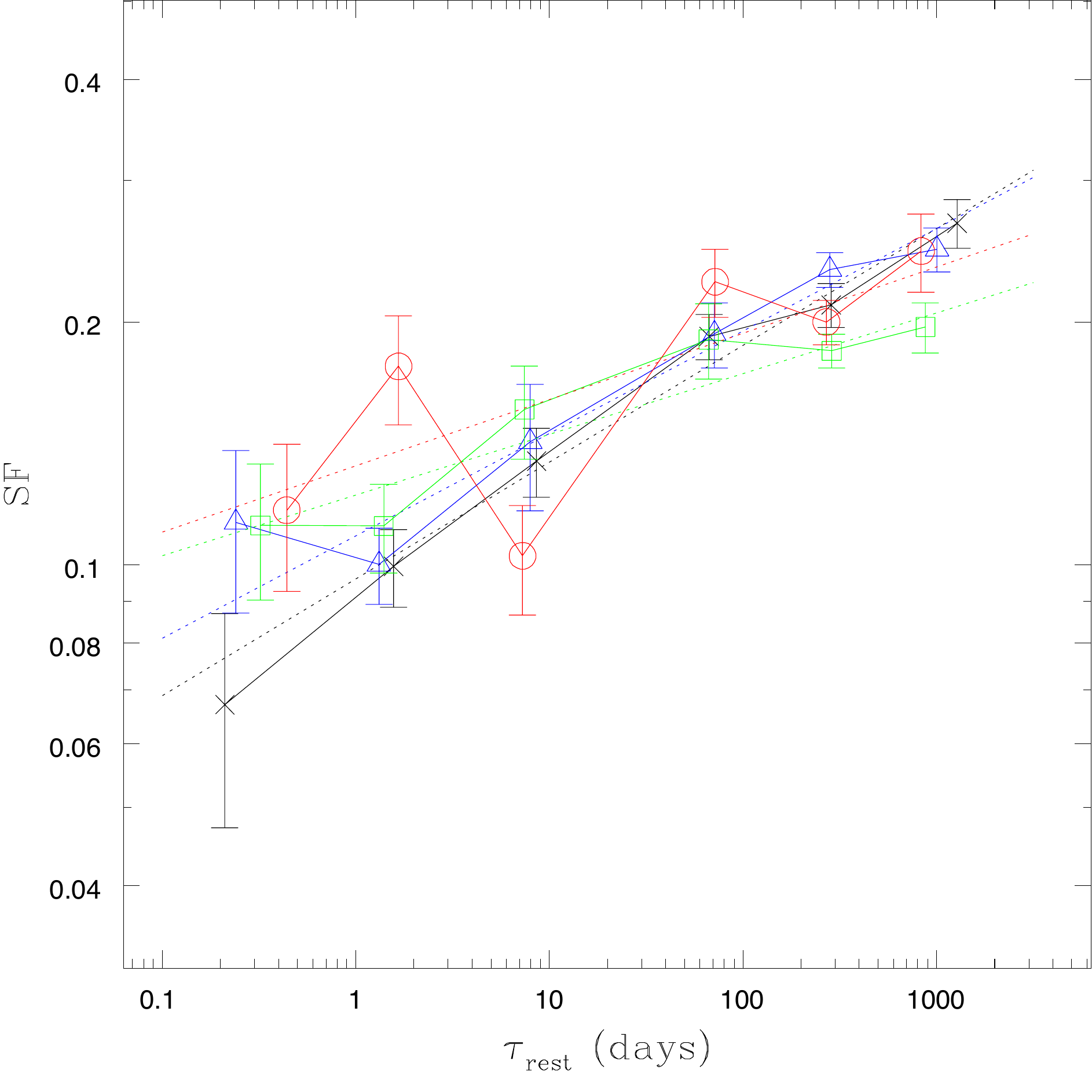}}
\caption{V-corrected SF in bins of redshift. Symbols and colours as in Fig. 11.}
\end{figure}

We also note that the previously discussed dependence of the SF on redshift can be affected. The $z$ dependent V-corrections for the four redshift bins adopted in Fig. 11 are V-corr=0.040, 0.058, 0.074, 0.085, in order of increasing redshift. The effect, shown in Fig. 13, is relatively small and suggests a weak increase with $z$ for the variability at short time lags, $\lesssim$ few days. However, this is not strong evidence because of both the relatively poor sampling and high error contribution for the SF at these time lags; see Figs. 6 and 7.

\section{Discussion}

The normalised excess variance is popularly used as a variability estimator. In most cases the method is applied correctly, using monitoring time intervals of fixed duration, for AGN samples at low redshift \citep[e.g. as in][]{pont12}. But this estimator depends on the length of the time interval in the rest frame and is therefore affected also by the cosmological time dilation \citep[e.g.][]{gask81}. The method is sometimes used improperly, choosing non-uniform time intervals, and/or including high redshift sources \citep[e.g.][]{lanz14}, thereby underestimating their variability. A few other examples of  this include the work by \citet{la-f14}, which applies NXS to the same data as \citet{pont12} to derive a luminosity distance estimator, but envisages an extension of the study to higher redshift sources, where NXS would underestimate variability. The work by \citet{cart15} applies NXS to the Quest-La Silla variability survey, including high redshift AGNs, whose variability is therefore underestimated. However, their main implication is a trend indicating that high redshift and more variable AGNs tend to have redder colours and this trend would be reinforced taking the NXS underestimate  into
account.

To demonstrate the duration effect for the NXS estimates, we used a sample of AGNs with multi-epoch X-ray observations (MEXSAS) extracted from the fifth release of the XMM-Newton Serendipitous Source Catalogue  (XMMSSC-DR5). We have also shown that the effect can be corrected on the basis of the knowledge of the behaviour of variability that is gained from structure function studies; our correcting formula, Eq. 9, can be successfully applied to further NXS-based studies.

We have updated the analysis of the ensemble structure function, finding that X-ray variability is well described by a power-law function of the rest-frame time lag, increasing as $\tau^{0.12}$ and extending up to $\sim 2000$ days. We have also shown that X-ray variability is inversely correlated with X-ray luminosity, approximately as $L_X^{-0.19}$. This anti-correlation has been reported, usually at short timescales, by many authors (e.g. \citet{barr86}, \citet{lawr93} for low-$z$ AGNs, \citet{mann02}, \citet{papa08} for higher $z$) with variability approximately proportional to $L_X^{-0.3}$. At longer timescales, the analysis by \citet{mark04}, for local AGNs, indicates $F_{var}\propto L_X^{-0.13}$. One simple interpretation of the anti-correlation is the superposition of several independently flaring subunits \citep[e.g.][]{gree93,nand97,alma00}.

We also find a dependence of variability on the emission frequency approximately as $\nu^{-0.15}$. In turn, this dependency is related to the change of the photon index, indicating a softer when brighter spectral variability behaviour, which extends a trend previously found for Seyfert galaxies \citep{sobo09} to AGNs with higher redshifts and luminosities. Because of this dependence, variability in the rest frame differs from that estimated in the observer-frame bands; however the effect can be corrected and we propose a simple correction term called V-correction, resulting in a moderate shift upwards ($\lesssim 15\%$) for the structure function. The same correction, applied in different bins of redshift,  can affect the resulting $z$-dependence of variability, suggesting a weak increase with $z$ for the variability at short time lags.

We finally remark that the corrections proposed by \citet{alle13} on the NXS should also be taken into account in the case of sparse sampling and for a comparison with physical models.

\begin{acknowledgements}
We acknowledge funding from PRIN/MIUR-2010 award 2010NHBSBE. We thank Stefano Bianchi, Szymon Koz\l owski, Fabio La Franca, Francesco Tombesi, and Dario Trevese for useful discussions. This research has made use of data obtained from the 3XMM XMM-Newton serendipitous source catalogue compiled by the 10 institutes of the XMM-Newton Survey Science Centre selected by ESA. Funding for SDSS-III has been provided by the Alfred P. Sloan Foundation, the Participating Institutions, the National Science Foundation, and the U.S. Department of Energy Office of Science. The SDSS-III web site is http://www.sdss3.org/. SDSS-III is managed by the Astrophysical Research Consortium for the Participating Institutions of the SDSS-III Collaboration including the University of Arizona, the Brazilian Participation Group, Brookhaven National Laboratory, Carnegie Mellon University, University of Florida, the French Participation Group, the German Participation Group, Harvard University, the Instituto de Astrofisica de Canarias, the Michigan State/Notre Dame/JINA Participation Group, Johns Hopkins University, Lawrence Berkeley National Laboratory, Max Planck Institute for Astrophysics, Max Planck Institute for Extraterrestrial Physics, New Mexico State University, New York University, Ohio State University, Pennsylvania State University, University of Portsmouth, Princeton University, the Spanish Participation Group, University of Tokyo, University of Utah, Vanderbilt University, University of Virginia, University of Washington, and Yale University.
\end{acknowledgements}

\bibliographystyle{aa}
\bibliography{nxs.bib}{}
\end{document}